# Method of diagnosing heart disease based on deep learning ECG sign

Jie Zhang[1,#], Lian Zou[1,*], Bohao Li[1], Kexin Xiang[1], Xuegang Shi[1], Cien Fan[1] and Yifeng Liu[2,*]

*[1] School of Electronic Information, Wuhan University*

*Wuhan, 430072, China*

*[2] China Academy of Electronics and Information Technology,*

*Beijing 100041, China*

*\*zoulian@whu.edu.cn*

*#2016301200121@whu.edu.cn*

The state-of-the-art method to classify heart rhythms by electrocardiogram (ECG) is to combine cardiology and signal processing method. The classification accuracy has been improved using deep learning models, but it still can't meet clinical needs. In this work, we propose a novel method to predict heart diseases from ECG signals with cardiology, signal processing methods and deep learning model. Firstly, we transform ECG signal into time-frequency diagram with wavelet transformation and cardiology. Then time-frequency diagram is then classified via a deep convolution network. Combing the advantages of signal processing method and deep learning methods, our method greatly improves classification accuracy. Extensive experiments show that our method achieved 0.87 F1-score (outperforming the best baseline by 0.11) in the 2017 PhysioNet/Computing-in-Cardiology Challenge for arrhythmia detection from single lead ECG classification. Our method also achieves state-of-the art result on different data sets, which demonstrates its robustness.

## 1. Introduction

Cardiovascular disease is the leading cause of death world-wide and the electrocardiogram (ECG) is a major tool in the diagnose[26]. Compared to determining whether ECG is normal or not, diagnosing heart disease is a more challenging task. Traditional ECG signal analysis and classification relies on the guidance of experienced medical experts, which is a time-consuming and labor-consuming job. Therefore, there is a growing need for fully automated ECG analysis using computer.

The difficulty in analyzing ECG is feature extraction. Current classification systems extract medical feature by signal processing means[2 3 4 5 8]. To classify the ECG, systems integrate the features and compare them with features extracted from various heart disease. However, due to noise interference, some features are difficult to extract[12]. Because ECG signals of different Cardiovascular disease have different features, it is impossible to design a system that can extract all the required features. This makes each diagnostic system poor scalability and low accuracy.

Recently, the deep learning technique has emerged as a promising solution for ECG interpretation[1 6 9 25]. Deep convolutional neural networks extract features automatically from the original images / signals[23 18]. Due to its high performance in automated classification, deep neural networks based ECG interpretation becomes progressively promising. For example, Pranav Rajpurkar[16] used a deep neural network to detect arrhythmias by the method of training a 34-layer convolutional neural network (CNN). Antônio H. Ribeiro and his research team used convolutional neural network a model for predicting electrocardiogram (ECG) abnormalities in short duration 12-lead ECG signals[17].

Although the two approaches mentioned above have their own advantages, the practical computer interpreted ECGs in clinical use have limitations in diagnostic accuracy and require human review and secondary interpretation. Given this, one might expect a complex method is required to achieve good results[19]. However, we show that a surprisingly simple and fast system can surpass prior state-of-the-art ECG classification results.

To solve the above problems, we propose a novel method to predict heart diseases from ECG signals with cardiology, signal processing methods and deep learning model. As shown in Figure1, the proposed system composes of two main parts: pre-processing module and a neural network architecture to classify heart rhythms. The pre-processing module first extracts feature waves of ECG signals using cardiology knowledge and signal processing methods. Then through the wavelet transformation, the 1-D feature waves are converted into 2-D time-frequency diagrams. In fact, the time-frequency diagrams are coefficient matrix of wavelet transformations at different scales. In this way, we can simultaneously analyze the time-frequency features of signals at different scales. The deep neural network extracts deep features of time-frequency diagrams and predict heart disease. Our consolidated contribution is three-fold.

1) we extract feature waves using cardiology knowledge and signal processing method. This method reduces the amount of computation while preserving the signal characteristics, which lead to better accuracy and faster execution.
2) we use wavelet transformation to convert 1-D signals into 2-D images, which allows deep learning model to simultaneously analyze the characteristics of different frequency bands of the signal[5 11 13].
3) we take full advantage of cardiology, signal processing methods and deep learning model, which demonstrated the robustness and achieved state-of-the art result.

The structure of this paper is as follows: in Section 2, we will focus on introducing our basic theory; in Section 3, we will discuss our experimental results. Finally, we will summarize our method in Section 4.

## 2. Methodology

Fig.1 illustrates our proposed system architecture. Since the variety of noise types may interfere the classification results, we preprocess the noise to improve accuracy and stability. First the raw ECG signals are filtered. Then, we extract R-waves in ECG using PT algorithm. According to the cardiology knowledge that the number of R-waves should be within a certain range, abnormal signals are replaced with zero sequences. This method can reduce the interference of noise to the classification of normal signals. Then, we extract a sequence containing four complete cardio cycles from each signal as feature wave. Afterwards, wavelet transformations of different scales are used to convert feature waves into time-frequency diagrams. Time-frequency diagrams are classified by convolutional neural networks.

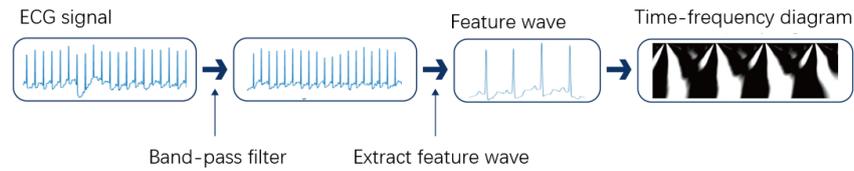

(a)

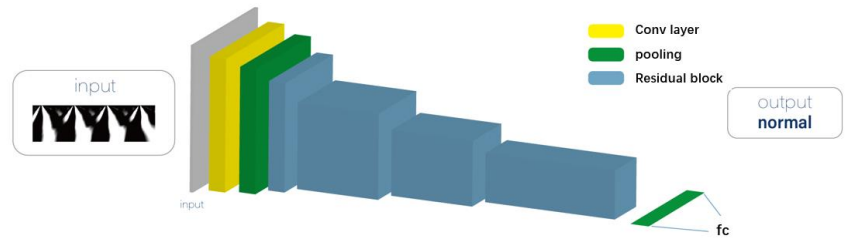

(b)

Fig.1. Overall frame diagram. (a) Time-frequency diagram conversion.
(b) the framework of classification network.

### 2.1 *Noise removal filtering*

Our system classifies heartbeat from single-lead electrocardiogram (ECG), which is a very weak low-frequency physiological signal. The movement of the electrodes, poor contact between the electrodes and the skin or other physiological electrical signals may interfere with the measured ECG signals, that causes the raw ECG signals to contain noise signals[21][22]. The frequency of ECG signals is between 0.5-100Hz and mainly concentrated between 3~35Hz[21]. Therefore, based on our study of noise signals, we designed a six-order Butterworth low-pass filter to filter out the high-frequency noise of the original ECG signal.

### 2.2 *Feature wave extraction*

Since the length of original ECG signals is a little long at the sampling frequency of 200Hz and not the same, it is very difficult to process the signals further. Besides, the diversity of noise signals may interfere with the signal classification. Therefore, we designed a method to extract feature waves of each ECG that can preserve signal characteristics while reducing interference of noise.

**2.2.1 *R wave extraction***

A normal ECG signal contains P, Q, R, S, T waves and heart disease may cause some waves to disappear in ECG[15]. For example, Atrial fibrillation (AF) may cause the disappearance of P waves. However, we find out that heart diseases generally do not result in the disappearance of R waves[20]. Therefore, we determine whether the signal is noise or not according to the number of the R waves.

We extract R waves with the modified Pan-Tompkins algorithm[14]. First an integer-coefficient band-pass filter is composed of cascaded low-pass and high-pass filters. Its function is noise rejection. Next a filter approximates a derivative. After an amplitude squaring process, the signal passes through a moving-window integrator. Adaptive thresholds then discriminate the locations of the QRS complexes.

The band-pass filter reduces the influence of muscle noise, 60 Hz interference, baseline wander, and T-wave interference. The desirable passband to maximize the QRS energy is approximately 5-15 Hz. For our chosen sample rate, we could not design a band-pass filter directly for the desired passband of 5-15 Hz using this specialized design technique. Therefore, we cascaded the low-pass and high-pass filters described below to achieve a 3 dB passband from about 5-12 Hz, reasonably close to the design goal.

The transfer function of the second-order low-pass filter is:

$$H_{lp}(z) = \frac{(1-z^{-6})^2}{(1-z^{-1})^2} \quad (1)$$

The amplitude response is:

$$|H(\omega T)| = \frac{sin^2(3\omega T)}{sin^2(\omega T/2)} \quad (2)$$

Where T=200 is the sampling period.

The design of the high-pass filter is based on subtracting the output of a first-order low-pass filter from an all-pass filter. The transfer function for such a high-pass filter is:

$$H(z) = \frac{(-1 + 32z^{-16} + z^{-32})}{(1+z^{-1})} \quad (3)$$

The amplitude response is:

$$|H(\omega T)| = \frac{[256 + sin^2(16\omega T)]^{1/2}}{cos(\omega T/2)} \quad (4)$$

After filtering, the signal is differentiated to provide the QRS complex slope information. We use a five-point derivative with the transfer function:

$$H(z) = (1/8T)(-z^{-2} - 2z^{-1} + 2z^1 + z^2) \quad (5)$$

To ensure that each sample is positive and nonlinear amplification of the output of the derivative emphasizes the higher frequencies, the signal is squared point by point. The equation of this operation is:

$$y(nT) = [x(nT)]^2 \qquad (6)$$

Then we use moving-window integration to obtain waveform feature information in addition to the slope of the R wave. It is calculated from:

$$y(nT) = (1/N)[x(nT - (N-1)T) + x(nT - (N-2)T) + \cdots + x(nT)] \qquad (7)$$

Where N=30 is the number of samples in the width of the integration window. If N is too large, the wave group will be flooded and if N is too small, other clutters will be generated.

Finally, Adaptive thresholds discriminate the locations of the R waves. Low thresholds are possible because of the improvement of the signal-to-noise ratio by the band-pass filter.

### 2.2.2 *Feature wave interception*

Based on cardiology, the number of R-waves in a fixed ECG signal should be within a certain range. If the total number of R waves is more or less than the range of values, it will be considered as a noise signal. For the stability of the system, we replace the original signals with fixed zero sequences as the feature waves of the noise signals. For ECG signals, irregular vibration may appear in the early stage due to adjustment of the acquisition equipment and the patient's posture. These signals cannot reflect the characteristics of the whole ECG signal. Besides, when doctors classify heart rhythms by ECG, they need to observe ECG of several cardiac cycle. Therefore, we proposed a method to extract feature waves[10][26]. We intercept feature waves containing four cardio cycles from the middle of ECG signals. Our experience shows that this method will reduce computational complexity while improving classification accuracy.

### 2.3 *Time-frequency diagram conversion*

In the field of signal processing, the time-frequency characteristics of signals can be obtained through wavelet transformation at different scales[13][10]. The wavelet transform formula is as follows:

$$W_f(a,b) = <f, \psi_{a,b}> = \frac{1}{\sqrt{a}} \int_{k\Delta t}^{(k+1)\Delta t} f(t) \times |a|^{-\frac{1}{2}} \varphi\left(\frac{t-b}{a}\right) \Delta t \qquad (8)$$

where $<*,*>$ is expressed as dot product, $a$ is expressed as scale factor, $b$ is expressed as displacement factor, $*$ is expressed as complex conjugate and $\psi_{a,b}(t)$ is expressed as wavelet basis function. In our method the wavelet basis function is db4 wavelet[25].

In order to apply the above formula to programming, we can achieve integration by discretizing the integral function and summing it. The discrete formula is as follows:

$$\begin{aligned} W_f(a,b) &= \sum_k \int_{k\Delta t}^{(k+1)\Delta t} f(t) \times |a|^{-\frac{1}{2}} \varphi\left(\frac{t-b}{a}\right) \Delta t \\ &= |a|^{-\frac{1}{2}} \Delta t \sum_k f(k\Delta t) \times \varphi\left(\frac{k\Delta t - b}{a}\right) \end{aligned} \qquad (9)$$

Using different wavelets for wavelet transformation can extract different time frequency characteristics of the signal. In order to analyze these features together, we proposed the following methods: we store wavelet transformation coefficients at different scales in a matrix orderly. The transformation coefficients for each scale are stored on each row of the matrix. When using convolution neural network to predict classification results, the coefficient matrix can be processed as a single-channel grayscale image. Examples of time-frequency diagrams corresponding to each kind of signal are shown in Table.1.

Table.1 the time-frequency diagrams of signal

| Class | Feature Wave | Time-Frequency Diagram |
|---|---|---|
| Normal | 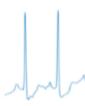 | 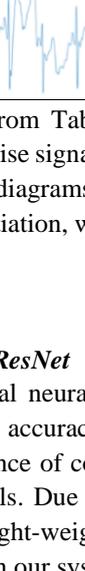 |
| AF | 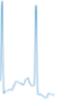 | 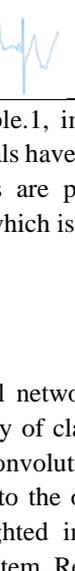 |
| Other Rhythm | 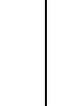 | 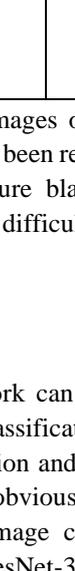 |
| Noise | 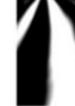 | 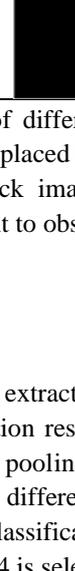 |

As can be seen from Table.1, images of different classes have different textures. Because most of the noise signals have been replaced by zero signals in the pre-processing, their time-frequency diagrams are pure black images. Images in other classes have similarity and differentiation, which is difficult to observe in 1-D ECG signals.

**2.4** *Classification via ResNet*

Deep convolutional neural network can extract deep features of images, which is helpful to improve the accuracy of classification results. For the capability to geometric transformation invariance of convolution and pooling modules, our systems can be used without aligning signals. Due to the obvious differences in time-frequency diagrams of different classes, a light-weighted image classification network can achieve a good classification results. In our system, ResNet-34 is selected as the classification network.

**3. Experimental result**

We used 8528 single lead episodic ECG records in the dataset of 2017 PhysioNet/Computing-in-Cardiology Challenge. Each ECG was annotated by cardiologists into four classes: normal, Atrial fibrillation (AF), other rhythm and noise.

We test our system on the testing dataset. The confusion matrix for the model predictions is shown in Fig.2. Most of arrhythmias can be detected with a proper high accuracy. Precision and recall of each class are shown in Table.2. The noise signals interfere classification results. However, noise interference can be avoided if ECG signals are properly collected in application. When we classify the signals after removing the noise signals, the classification accuracy can reach more than 98.6% on average.

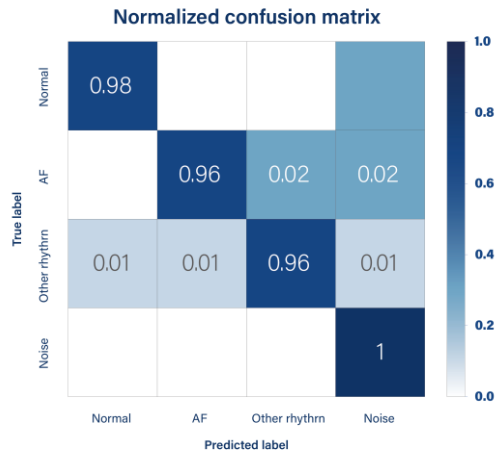

Fig. 2. the classification results of using resnet34

Table 2: the results on the test set

| Class | Precision | Recall |
|---|---|---|
| Normal | 99.32% | 96.67% |
| AF | 97.83% | 90.00% |
| Other rhythm | 98.44% | 90.00% |
| Noise | 68.18% | 100% |

According to the ranking method of competition, we calculated the F1 score of our system. As we can see in Table.3, the F1 score of our system is 0.11 higher than the winner of the competition. Our system is a surprisingly simple and fast, which can surpass prior state-of-the-art ECG classification results. Besides, our method also shows robust performance and achieved state-of-the-art result on different data sets.

Table 3: Comparison with the first few

| | Participant | F1 score |
|---|---|---|
| 1st | Guangyu Bin | 0.86 |
| 2nd | Zhaohanx | 0.85 |

| | Tomas.teijeiro | 0.85 |
| --- | --- | --- |
| | Fplesinger | 0.85 |
| 5th | Rmaka08 | 0.84 |
| | **Our method** | **0.87** |

## 4. Conclusion

In this work, we propose a novel method to predict heart diseases from ECG signals with cardiology, signal processing methods and deep learning model. We use wavelet transformation to convert 1-D signals into 2-D images, which allows deep learning model to simultaneously analyze the characteristics of different frequency bands of the signal. Our system is simple, fast and shows the state-of-the-art performance.

Due to the excellent performance of our system in the task of four classifications of ECG signals, we expect this method to apply to more complex tasks. We will experiment on larger data sets with finer annotations. In addition, since different wavelets can extract different features of frequency when using wavelet transform, the performance of our system is different. We will try to use other wavelets to process the signals.

## Acknowledgements

This work was supported by the CETC Foundation (No. 6141B0801010a, No. 6141B08010102 and No. 6141B08080101)